# Comment on
# "The effect of variable viscosity on mixed convection heat transfer along a vertical moving surface" by M. Ali [International Journal of Thermal Sciences, 2006, Vol. 45, pp. 60-69]

Asterios Pantokratoras
Associate Professor of Fluid Mechanics
School of Engineering, Democritus University of Thrace,
67100 Xanthi – Greece
e-mail:apantokr@civil.duth.gr

## 1. INTRODUCTION

The problem of forced convection along an isothermal moving plate is a classical problem of fluid mechanics that has been solved for the first time in 1961 by Sakiadis (1961). It appears that the first work concerning mixed convection along a moving plate is that of Moutsoglou and Chen (1980). Thereafter, many solutions have been obtained for different aspects of this class of boundary layer problems. In the previous works the fluid properties have been assumed constant. Ali (2006) in a recent paper treated, for the first time, the mixed convection problem with variable viscosity. He used the local similarity method to solve this problem but there are doubts about the validity of his results. For that reason we resolved the above problem with the direct numerical solution of the boundary layer equations without any transformation.

## 2. THE MATHEMATICAL MODEL

Consider the flow along a vertical flat plate with u and v denoting respectively the velocity components in the x and y direction, where x is the coordinate along the plate and y is the coordinate perpendicular to x. For steady, two-dimensional flow the boundary layer equations including variable viscosity and buoyancy forces are

continuity equation: $$\frac{\partial u}{\partial x} + \frac{\partial v}{\partial y} = 0 \qquad (1)$$

momentum equation: $$u\frac{\partial u}{\partial x} + v\frac{\partial v}{\partial y} = \frac{1}{\rho_a}\frac{\partial}{\partial y}\left(\mu\frac{\partial u}{\partial y}\right) + g\beta(T - T_a) \qquad (2)$$

energy equation:
$$u\frac{\partial T}{\partial x} + v\frac{\partial T}{\partial y} = \alpha \frac{\partial^2 T}{\partial y^2} \qquad (3)$$

where T is the fluid temperature, μ is the dynamic viscosity, α is the thermal diffusivity, and $\rho_a$ is the ambient fluid density.
The following boundary conditions have been applied:

$$\text{at } y = 0 \quad u=U_w, v=0, T=T_w \qquad (4)$$
$$\text{as } y \rightarrow \infty \quad u=0, T=T_a \qquad (5)$$

where $T_w$ is the plate temperature, $T_a$ is the ambient fluid temperature and $U_w$ is the constant velocity of the moving plate.

The viscosity is assumed to be an inverse linear function of temperature given by the following equation (Ali 2006)

$$\frac{1}{\mu} = \frac{1}{\mu_a}[1 + \gamma(T - T_a)] \qquad (6)$$

where $\mu_a$ is the ambient fluid dynamic viscosity and γ is a thermal property of the fluid.

The equations (1)-(3) represent a two-dimensional parabolic flow. Such a flow has a predominant velocity in the streamwise coordinate (unidirectional flow) which in our case is the direction along the plate. The equations were solved directly, without any transformation, using the finite difference method of Patankar (1980). The solution procedure starts with a known distribution of velocity and temperature at the plate edge (x=0) and marches along the plate. At the leading edge the temperature was taken uniform and equal to ambient one and the velocity was also uniform with a very small value. At each downstream position the discretized equations (2) and (3) are solved using the tridiagonal matrix algorithm (TDMA). The cross-stream velocities v were obtained from the continuity equation. The forward step size Δx was 0.001 mm and we used a nonuniform lateral grid with 500 points where Δy increases along y. In the numerical solution of the boundary layer problems the calculation domain must always be at least equal or wider than the boundary layer thickness. However, it is known that the boundary layer thickness increases with x. Therefore, it would be desirable to have a grid which conforms to the actual shape of the boundary layer. For that reason an expanding grid has been used in the present work. The results are grid independent. The parabolic solution procedure is a well known solution method and has been used extensively in the literature. It appeared for the first time in 1970 (Patankar and Spalding, 1970) and has been included in classical fluid mechanics textbooks (see page 275 in White, 1991). Anderson et al. (1984) mention 7 numerical methods for the solution of the boundary layer equations (page 364) and among them is the "well known Patankar–Spalding method". The method is fully implicit and can be applied to both similar and nonsimilar problems. The dynamic viscosity µ and the Prandtl number, which is a function of viscosity, have been considered variable during the solution procedure. A detailed description of the solution procedure, with variable thermophysical properties, may be found in Pantokratoras (2002).

## 3. RESULTS AND DISCUSSION

The local Nusselt number and the local Reynolds number have been defined as follows by Ali (2006)

$$Nu_x = \frac{hx}{k} \tag{7}$$

$$Re_x = \frac{U_w x}{\nu_a} \tag{8}$$

thus the term $Nu_x Re_x^{-0.5}$ is

$$Nu_x Re_x^{-0.5} = \frac{hx}{k} Re_x^{-0.5} = -\frac{xk}{k(T_w - T_a)} Re_x^{-0.5} \left[\frac{\partial T}{\partial y}\right]_{y=0} = -\frac{x}{T_w - T_a} Re_x^{-0.5} \left[\frac{\partial T}{\partial y}\right]_{y=0} \tag{9}$$

The quantity $C_f$ has not been defined by Ali (2006) and we used the following equation for this quantity (Bejan 1995, page 51)

$$C_f = \frac{2\tau_w}{\rho_a U_w^2} \tag{10}$$

where $\tau_w$ is the wall shear stress given by

$$\tau_w = \mu_w \left[\frac{\partial u}{\partial y}\right]_{y=0} \tag{11}$$

Consequently the term $C_f Re_x^{0.5}$ is

$$C_f Re_x^{0.5} = \frac{2\mu_w}{\rho_a U_w^2} Re_x^{0.5} \left[\frac{\partial u}{\partial y}\right]_{y=0} \tag{12}$$

Ali (2006) transformed equations (1)-(3) into the following equations

$$f''' - \frac{\theta - \theta_r}{\theta_r} ff'' - \frac{1}{\theta - \theta_r}\theta' f'' - \frac{2\lambda(\theta - \theta_r)}{\theta_r}\theta = 0 \tag{13}$$

$$\theta'' + \Pr f\theta' = 0 \tag{14}$$

where f and θ are the dimensionless velocity and dimensionless temperature defined as

$$f' = \frac{u}{U_w} \tag{15}$$

$$\theta = \frac{T - T_a}{T_w - T_a} \tag{16}$$

$\lambda = Gr_x/Re_x^2$ is the buoyancy parameter and $Gr_x$ is the Grashof number defined as

$$Gr_x = g\beta(T_w - T_a)x^3/\nu_a^2 \tag{17}$$

$\theta_r$ is the viscosity parameter defined by

$$\theta_r = -\frac{1}{\gamma(T_w - T_a)} \tag{18}$$

It should be mentioned here that when $\theta_r \to \infty$ the fluid viscosity becomes equal to ambient viscosity. In equations (13) and (14) the prime represents differentiation with respect to similarity variable η defined as (Ali, 2006)

$$\eta = \frac{y}{x\sqrt{2}} Re_x^{1/2} \tag{19}$$

Ali (2006) solved equations (13) and (14) using the fourth order Runge-Kutta method. Locally similarity solutions were obtained for increasing values of λ at each constant $\theta_r$. At each new $\theta_r$ the procedure starts from a known solution which corresponds to pure forced convection (λ=0). The Prandtl number included in the transformed energy equation (15) was assumed constant and equal to ambient Prandtl number

$$\Pr_a = \frac{\nu_a}{a} \tag{20}$$

However, the Prandtl number is a function of viscosity and as viscosity varies across the boundary layer, the Prandtl number varies, too.

In table 1 the skin friction coefficient $C_f Re_x^{0.5}$ and the Nusselt number $Nu_x Re_x^{-0.5}$ are given for ambient Prandtl number 0.72. In this table the results by Ali (2006) have been also included for comparison. The results by Ali have been taken from his figures 4 and 8. It was difficult to extract values for $\theta_r$ near 0 and 1 and for that reason we took values for $-10 \leq \theta_r \leq -1.0$ and $1.5 \leq \theta_r \leq 10$. In the last column of the table the Prandtl numbers at the plate ($Pr_w$) are included.

$$Pr_w = \frac{v_w}{a} \qquad (21)$$

Table 1. Values of $C_f Re_x^{0.5}$ and $Nu_x Re_x^{-0.5}$ for $Pr_a = 0.72$

| $\theta_r$ | $C_f Re_x^{0.5}$ | | | $Nu_x Re_x^{-0.5}$ | | | $Pr_w$ |
| | Present Work | Ali (2006) | Difference % | Present Work | Ali. (2005) | Difference % | |
|---|---|---|---|---|---|---|---|
| $\lambda=0$ | | | | | | | |
| ∞ constant viscosity | -0.8854 (-0.8875 from Moutsoglou and Chen for Pr=0.7) | -0.88 | <1 | 0.3555 (0.3492 from Moutsoglou and Chen for Pr=0.7) | 0.35 | <1 | 0.72 |
| $\lambda=1$ | | | | | | | |
| ∞ constant viscosity | 0.3886 (0.3885 from Moutsoglou and Chen for Pr=0.7) | 0.88 | 126 | 0.4559 (0.4550 from Moutsoglou and Chen for Pr=0.7) | 0.46 | <1 | 0.72 |
| -10 | 0.3834 | 0.88 | 129 | 0.4571 | 0.46 | <1 | 0.65 |
| -7.5 | 0.3866 | 0.88 | 128 | 0.4588 | 0.46 | <1 | 0.64 |
| -5.0 | 0.3839 | 0.88 | 129 | 0.4590 | 0.46 | <1 | 0.60 |
| -2.5 | 0.3846 | 0.88 | 129 | 0.4591 | 0.46 | <1 | 0.51 |
| -1.0 | 0.3850 | 0.88 | 129 | 0.4620 | 0.46 | <1 | 0.36 |
| 1.5 | 0.4602 | 0.88 | 91 | 0.4524 | 0.46 | <1 | 2.16 |
| 2.5 | 0.4010 | 0.88 | 119 | 0.4537 | 0.46 | <1 | 1.20 |
| 5.0 | 0.3898 | 0.88 | 126 | 0.4551 | 0.46 | <1 | 0.90 |
| 7.5 | 0.3894 | 0.88 | 126 | 0.4563 | 0.46 | <1 | 0.83 |
| 10.0 | 0.3891 | 0.88 | 126 | 0.4569 | 0.46 | <1 | 0.80 |

| | λ=5 | | | | | | |
|---|---|---|---|---|---|---|---|
| ∝ constant viscosity | 4.2621 (4.2798 from Moutsoglou and Chen for Pr=0.7) | 5.71 | 34 | 0.5987 (0.5909 from Moutsoglou and Chen for Pr=0.7) | 0.59 | <1 | 0.72 |
| -10 | 4.1356 | 5.71 | 38 | 0.6020 | 0.59 | <1 | 0.65 |
| -7.5 | 4.1284 | 5.63 | 36 | 0.6022 | 0.59 | <1 | 0.64 |
| -5.0 | 4.0488 | 5.58 | 38 | 0.6063 | 0.60 | <1 | 0.60 |
| -2.5 | 3.8859 | 5.45 | 40 | 0.6153 | 0.61 | <1 | 0.51 |
| -1.0 | 3.4497 | 5.00 | 45 | 0.6267 | 0.62 | <1 | 0.36 |
| 1.5 | 5.5848 | 7.62 | 36 | 0.5593 | 0.55 | <1 | 2.16 |
| 2.5 | 4.9147 | 6.59 | 34 | 0.5775 | 0.57 | <1 | 1.20 |
| 5.0 | 4.5692 | 6.31 | 38 | 0.5879 | 0.58 | <1 | 0.90 |
| 7.5 | 4.4235 | 5.98 | 35 | 0.5932 | 0.59 | <1 | 0.83 |
| 10.0 | 4.4102 | 5.98 | 36 | 0.5965 | 0.59 | <1 | 0.80 |
| | λ=10 | | | | | | |
| ∝ constant viscosity | 8.2504 (8.29 from Chen for Pr=0.7) | 10.87 | 32 | 0.6884 (0.6800 from Chen for Pr=0.7) | 0.68 | <1 | 0.72 |
| | λ=20 | | | | | | |
| -10 | 14.8304 | 18.94 | 28 | 0.8054 | 0.79 | 2 | 0.65 |
| -7.5 | 14.7219 | 18.87 | 28 | 0.8085 | 0.79 | 2 | 0.64 |
| -5.0 | 14.3042 | 18.62 | 30 | 0.8159 | 0.80 | 2 | 0.60 |
| -2.5 | 13.6585 | 17.75 | 30 | 0.8295 | 0.81 | 2 | 0.51 |
| -1.0 | 12.0610 | 16.00 | 33 | 0.8574 | 0.84 | 2 | 0.36 |
| 1.5 | 20.2868 | 24.91 | 23 | 0.7297 | 0.69 | 5 | 2.16 |
| 2.5 | 17.6649 | 22.07 | 25 | 0.7795 | 0.75 | 4 | 1.20 |
| 5.0 | 16.2364 | 20.60 | 27 | 0.7873 | 0.76 | 3 | 0.90 |
| 7.5 | 15.8372 | 20.17 | 27 | 0.7901 | 0.77 | 3 | 0.83 |
| 10.0 | 15.6426 | 20.00 | 28 | 0.7920 | 0.78 | 2 | 0.80 |

From table 1 it is seen that the skin friction coefficient $C_f Re_x^{0.5}$ and the Nusselt number $Nu_x Re_x^{-0.5}$ calculated by the present method are in very good agreement with those calculated by Ali (2006) and Moutsoglou and Chen (1980) for the case λ=0 (pure forced convection) and constant viscosity. Except that the above quantities calculated by the present method are in very good agreement with those calculated by Moutsoglou and Chen (1980) for the cases λ=1, 5 (mixed convection) and constant viscosity. Our results compare also very well with those of Chen (2000) for λ=10 and constant viscosity. In addition our method has been used recently successfully to two similar problems (Pantokratoras, 2004, 2005). The Nusselt numbers given by Ali

(2006) are in good agreement with our results for all cases of the buoyancy parameter $\lambda$. For the skin friction coefficient $C_f Re_x^{0.5}$ things are different. For $\lambda=0$ there is very good agreement but for $\lambda=1$ large differences appear. The divergence exist also for higher values of $\lambda$ at a smaller rate. It is seen that our $C_f Re_x^{0.5}$ values are always lower than those of Ali (2006) and this is in accordance with the velocity profiles included in figures 1 and 2 where we see that the velocity profiles calculated by the present method lay lower than those of Ali. It is advocated here that the results of the skin friction coefficient given by Ali (2006) for $\lambda \geq 1$ are wrong. The error is caused probably by the local similarity method that has been used for the solution of the equations. Minkowycz and Sparrow (1974) mention that an unorthodox version of the local similarity method yields results of uncertain accuracy. It should be noted here that Ali (2006) tested the accuracy of his method comparing the results only with those of the pure forced convection case ($\lambda=0$). If the comparison had been extended to existing results for the mixed convection problem with constant viscosity (Moutsoglou and Chen, 1980, Chen, 2000) the error would appear.

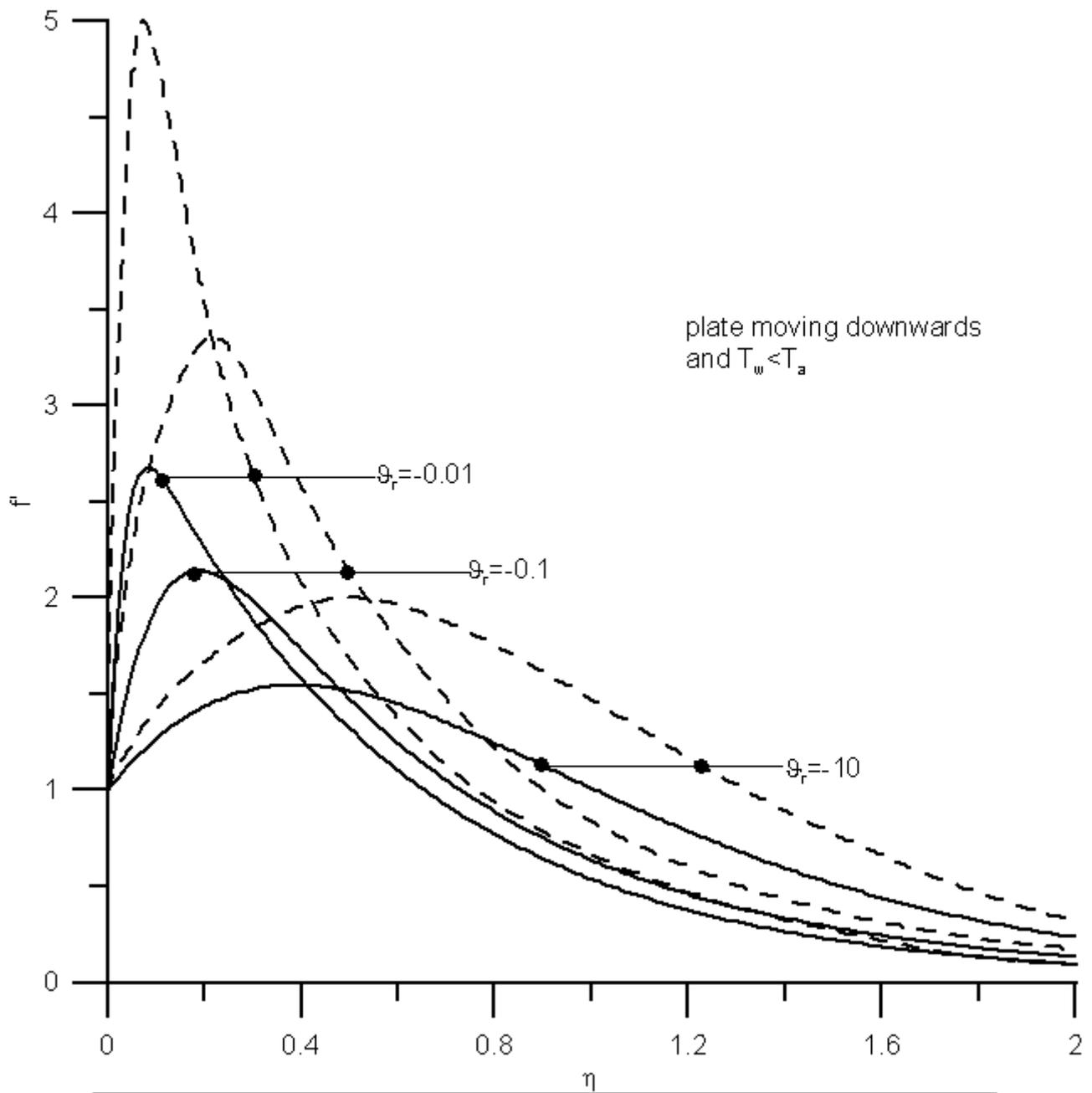

Figure 1. Velocity profiles for a downwards moving plate, Pr=0.72 and λ=5: Solid line, present work :dashed line, Ali(2006).

plate moving downwards and $T_w<T_a$

$\vartheta_r=-0.01$
$\vartheta_r=-0.1$
$\vartheta_r=-10$

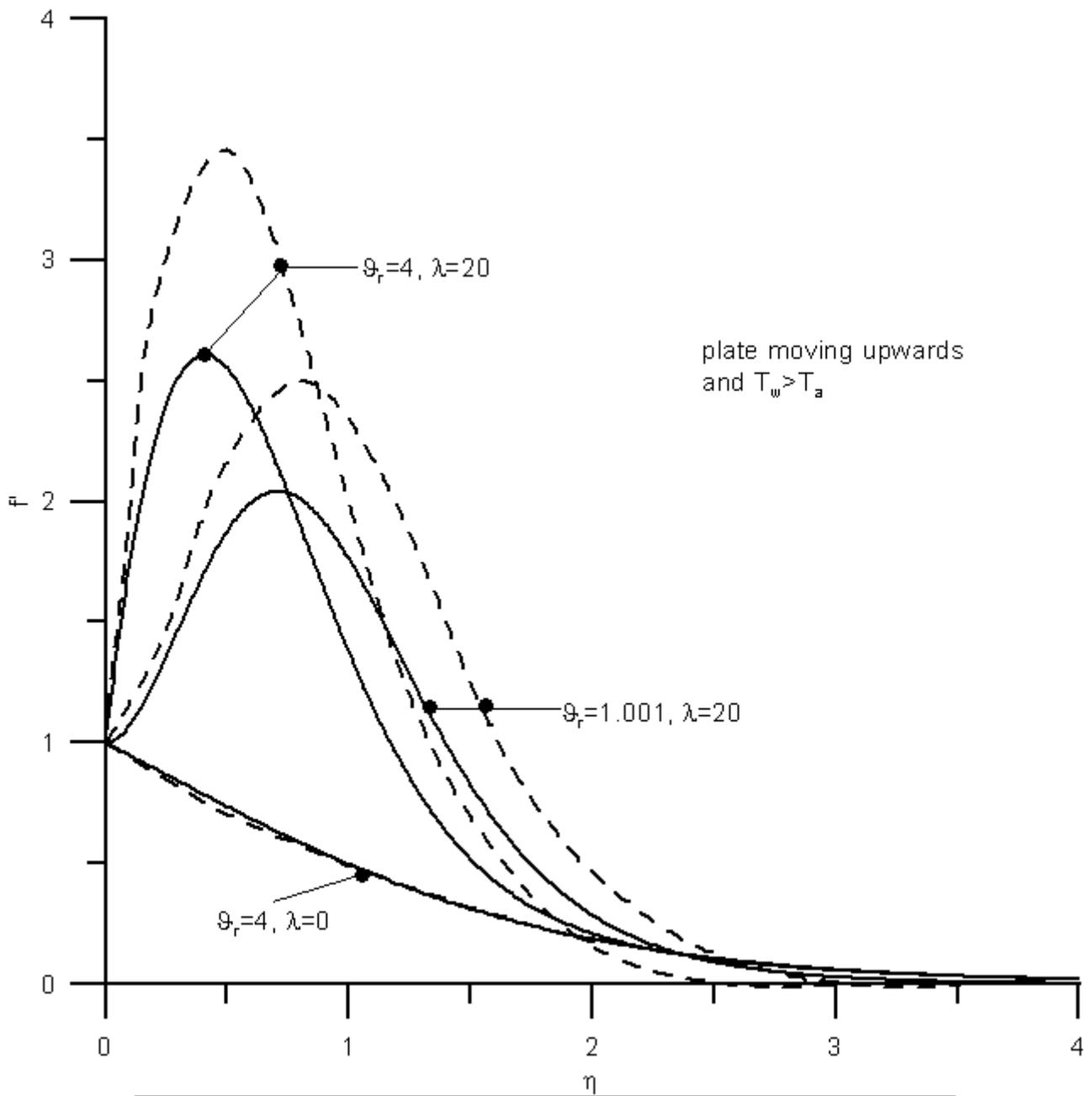

Figure 2. Velocity profiles for an upwards moving plate and Pr=0.72 : Solid line, present work: dashed line, Ali(2006).